\documentstyle[12pt,epsf]{article} 
\begin{document}
\baselineskip 18pt

\title{Criticality in driven cellular automata with defects}
\author{Bosiljka  Tadi\'c$^\diamond$  and Ramakrishna Ramaswamy$^*$\\
\it{$^\diamond$Jo\v zef Stefan Institute,  P.O. Box 100, 61111 Ljubljana, Slovenia}\\
\it{$^*$School of Physical Sciences,}\\ 
\it{Jawaharlal Nehru University, New Delhi 110067 India}}

\date{}

\maketitle

\begin{abstract}

We study three models of driven sandpile-type automata in the
presence of quenched random defects.  When the dynamics is
conservative, all these models, termed the random sites (A),
random bonds (B), and random slopes (C), self-organize into a critical
state.  For Model C the concentration-dependent exponents
are nonuniversal.  In the case of nonconservative defects, the
asymptotic state is subcritical.  Possible defect-mediated
nonequilibrium phase transitions are also discussed.

\end{abstract}

\section{Introduction}

Self-organized criticality (SOC)~\cite{BTW} is a dynamic phenomenon
which occurs in certain dissipative systems with large numbers of
degrees of freedom.  Such a system, when slowly driven into its
metastable state, self-organizes in a state with long-range
correlations, similar to the critical state at a
second-order phase transition.

In conventional criticality, quenched disorder can be a {\it relevant}
perturbation in the vicinity of the critical point.  It is therefore
natural to ask how the SOC state responds to similar perturbations.
One might expect that self-organizing systems are more robust
against random perturbations, since the critical state is an
attractor of the dynamics, although the universality class may
change in presence of disorder.

We explore this question in the present work, where we demonstrate,
using numerical simulations on simple models of self-organizing
cellular automata with frozen random defects, the conditions for a
system to self-organize, and determine the universality class of the
critical behavior.  Disorder-mediated phase transition between different
types of metastable states are also discussed. 

We study three kinds of random defects which locally affect the rules of
relaxation in a manner analogous to the random site, random bond, and
random field defects in spin models displaying conventional critical
phenomena.  All these models are based on the directed abelian
2-dimensional {\it critical height} model (toppling if $h(i,j) \ge h_c$ )
which is exactly solvable in the absence of defects~\cite{DR}, in which
the dynamic rules are locally modified at a fraction $c$ of ``defect''
sites.  In model A, holes of infinite depth are placed at random sites.
In model B there are two variables $h_1(i,j)$ and $h_2(i,j)$ associated
with each site $(i,j)$. These are coupled at random sites (details are
given in Section 3 below), in a manner so as to lead to a multiplicity of
metastable states, as in the case of spin-glasses and other frustrated
systems.  In model C, frozen-in local slopes are introduced by preventing
relaxation through the height instability at defect sites, and instead
applying a critical slope toppling rule at all sites with slopes $\sigma
(i,j) \ge \sigma _c$.  In this case, regions with local slopes $\sigma
(i,j) \sim \sigma _c$ are established, while the rest of the system
relaxes according to the critical height rule (the critical height $ h_c$
is chosen such that $ h_c \mathopen< \sigma _c$).  In all three cases the
ratio between the number of particles leaving one site and the number
appearing at its neighbors is not uniform at defect sites.

The numerical results presented here are from simulations on lattices
(with periodic boundary conditions in the transverse direction) of size
$12 \le L \le 384$ for time steps (see below) up to $1 \times 10^6$.  A
lattice with frozen-in defects is prepared and kept fixed for the entire
number of time steps, and then a new configuration is prepared; results
are then averaged over the total number of configurations. In each case,
the system was driven by randomly adding particles to sites at the top
(first row), $h(1,j) \to h(1,j)+1$.

\section{Subcriticality: random-site model (A)}

The dynamic height variable $h(i,j)$, associated with each lattice
site $(i,j)$ on a square lattice, is updated according to the rules 
of the critical height model: if 
$h(i,j)$ exceeds a critical value $h_c$, then the site is
unstable and relaxes according to

\begin{equation} h(i,j) \to h(i,j)-2 \; ~~~~;~~~~h(i+1,j_\pm) \to
h(i+1,j_\pm)+1 \; , \label{CHR} \end{equation} 

where $(i+1,j_\pm)$ are the two neighboring downstream sites.  This
rule applies to all sites except for a fraction $c$ of randomly
distributed defect sites, at which infinitely deep holes are placed.
Thus two grains are lost each time an avalanche hits a defect,
rendering the dynamic process nonconservative. (Similar nonconservative
models have been considered earlier \cite{MKK}).  For $c = 0$ the
critical exponents are listed in Table~1.  An annealed version of
defects of the above type was studied numerically in Ref.\
\cite{TNURP}, where it was shown that the system self-organizes into
a subcritical state.  The probability distributions of duration
$P(t,c)$ and size $D(s,c)$ were found to fulfill the following
scaling forms

\begin{equation} 
P(t,c) = t^{-\alpha }{\cal{P}}(t/\xi _t(c)) \; ~~~~
;~~~~ D(s,c) = s^{-\tau }{\cal{D}}(s/\xi _s(c)) \ , 
\label{rchsc}
\end{equation} 
where the correlation lengths $\xi _t(c) \sim c^{-\mu _t}$ and $\xi _s(c)
\sim c^{-\mu _s}$ for time and spatial correlations,
respectively, diverge in the limit $c \to 0$ with $\mu _t =1$ and $\mu
_s=1.5$~\cite{TNURP,Theiler}.  The scaling functions defined in Eq.\
(\ref{rchsc}) were calculated numerically in Ref.\ \cite{TNURP}.  With the
dynamical rules (\ref{CHR}) there are no recurrences in the avalanche
dynamics and thus both annealed and quenched (frozen) impurities lead to
the same universal scaling properties.  In Fig.\ 1 we plot $<n(l)>$, the
average number of particles relaxed \cite{dissip} up to distance $l$
(measured from the top of the pile) for various values of the defect
concentration:
\begin{figure}[htb] 
\epsfxsize=12cm 
\epsffile[0 100 507 405]{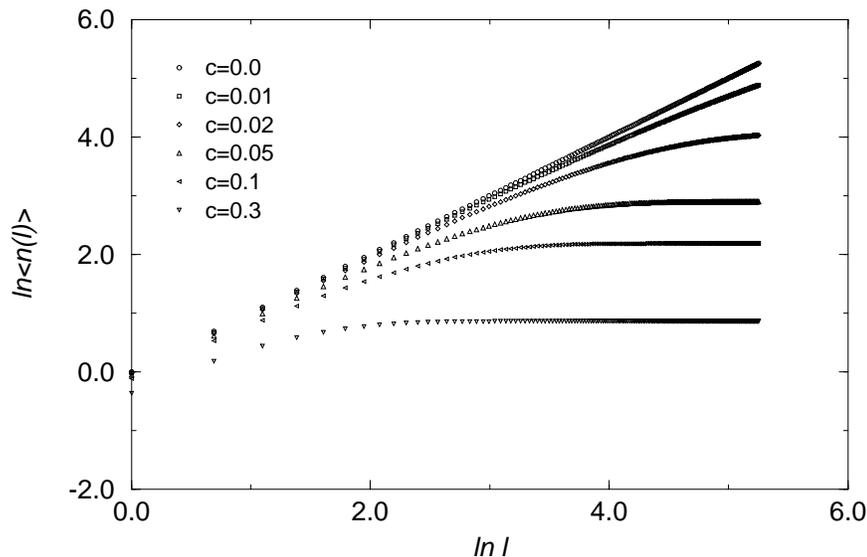} 
\label{fig1} 
\caption{Time average of the number
of topplings $<n(l)>$ vs.  distance $l$ from the top of
pile for various concentrations of defects $c$.}  
\end{figure}

In the absence of defects $<n(l)>$ is proportional to the average 
number of toppled sites $<s(l)>$ and thus exhibits a power-law behavior 
\begin{equation}
<n(l)> \sim l^\mu 
\label{newq}
\end{equation}
with $\mu = 1$ (cf.  the top curve in Fig.\ 1).  For finite concentration
of defects $c$, the curve flattens, and the scaling region with slope
$\mu$ decreases with increasing $c$, corresponding to the decreasing
correlation length in the system.  At a critical concentration $c = c^* =
0.295 \equiv 1-p_d$, the directed-percolation threshold~\cite{dirperc}, it
is no longer possible to have a lattice spanning cluster~\cite{TNURP}.

For sandpiles without defects, the fluxes into and out of the lattice
are equal-- the distribution $G(J,L)$ of the current which flows
over the rim of the system is shown in Fig.\ 2 for $c = 0.0$ and
various systems sizes $L$.  As expected for a self-organized
critical state, the outflow current distribution exhibits
scale invariance~\cite{Ketal}, {\it i.e.}

\begin{equation}
G(J,L) = L^{-\beta }{\cal{G}}(J/L^\phi ) \ . 
\label{jscal}
\end{equation}

\begin{figure}[htb] 
\epsfxsize=12cm 
\epsffile[20 70 520 360]{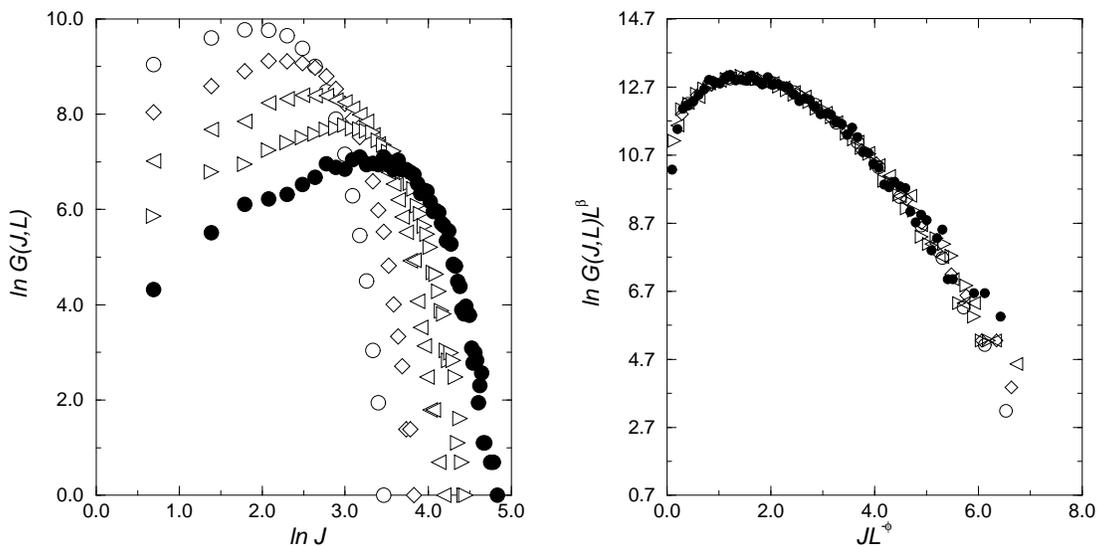} 
\label{fig2} 
\caption{Outflow current
distribution in the absence of defects for system sizes $L=$
24, 48, 96, 192, and 384 (left), and the corresponding finite-size
scaling plot (right).}  
\end{figure} 
The finite-size scaling fit, also shown in Fig.\ 2 has the
exponents $\beta =1$ and $\phi = 0.5$.  With defects, the outflow
current diminishes with increasing concentration of defects,
eventually vanishing at $c=c^*$ (see Section~5).  There is no apparent
scaling form of $G(J,c)$.

\section{Universal criticality: random-bond model (B)}

In order to simulate effects of random bonds in a sandpile
automaton, we introduce a two-state variable $\vec{h} = (h_1,h_2)$
at each lattice site $(i,j)$, where $h_1$ and $h_2$ are 
not necessarily integer.  Each bond carries a quenched variable 
$b$ with value $b=\pm 1$:  the
disorder here is that a random fraction $c$ of the bonds have $b =
-1$.  The evolution rule for model B depends on the absolute value
of the difference between components $h_1$ and $h_2$ at a site,
which causes instability if it exceeds a critical value $d_c$.  The
entire number of particles then topple, and the two downstream
neighboring sites are updated as follows (similar models were
discussed earlier in Ref.~\cite{MH}):

\begin{equation}
{\mathrm{If}}~~ \, |h_1(i,j)-h_2(i,j)| \ge d_c \, \label{RBC}
\end{equation}
then
\begin{equation}
h_1(i,j) \to 0 \; ~~~;~~~ h_2(i,j) \to 0 \, \label{RB2}
\end{equation}
and
\begin{equation}
h_1(i+1,j\pm) \to h_1(i+1,j\pm) +(\lambda h_1(i,j)+h_2(i,j))/2 \, ,
~~~~{\mathrm{if}}~~ b=+1 \; 
\end{equation}
\begin{equation}
h_2(i+1,j\pm) \to h_2(i+1,j\pm) +(\lambda h_2(i,j)+h_1(i,j))/2 \, ,
~~~~{\mathrm{if}}~~ b=-1 \; 
\label{RB3}
\end{equation}
For $\lambda =1$ the dynamics is conservative: all particles which
leave site $(i,j)$ appear at its neighbors.  The instability
condition  Eq.\ (\ref{RBC}) leads to a variety of states
with high local values of $h_1$ and $h_2$.  The asymptotic state is,
however, SOC, and it appears that the precise form of the coupling
between two states $h_1$ and $h_2$ is unimportant (see also Ref.\
\cite{Peng}). Such a model incorporates some features of
neural networks, and introduces frustration effects~\cite{MH}.

\begin{figure}[htb] 
\epsfxsize=12cm 
\epsffile[20 65 480 356]{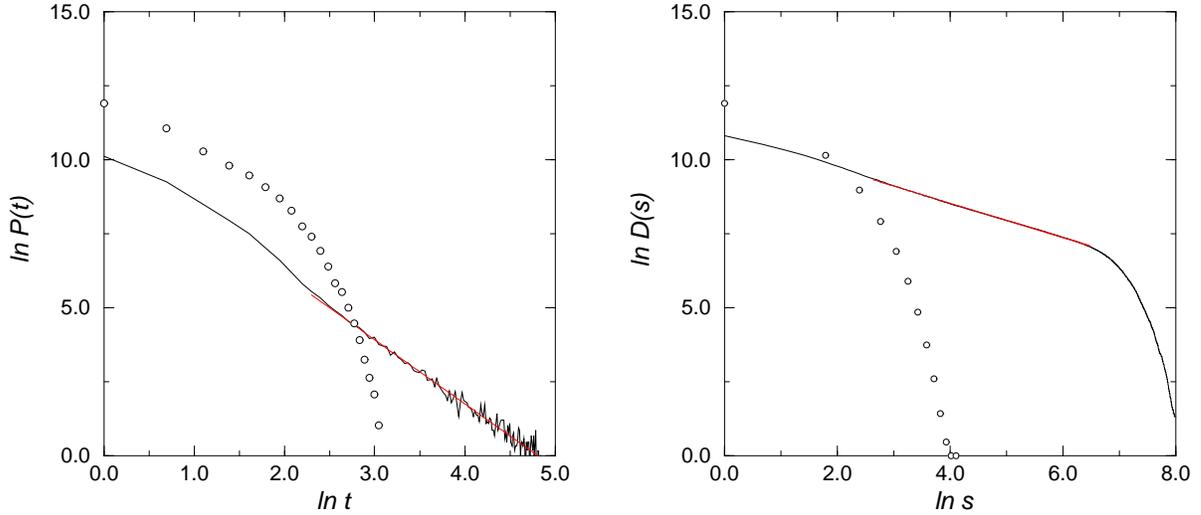} 
\label{fig3}
\caption{Distributions of duration $P(t)$ and size $D(s\ge s_0)$ of
the relaxation clusters in model B for conserving $\lambda =1$
(solid lines) and nonconserving dynamics $\lambda \mathopen< 1$
(open circles).}  
\end{figure}

It is necessary to have conservative dynamics ($\lambda =1$) in
order for the system to self-organize into a critical state.
Results of numerical simulations for the distributions of duration
$P(t)$ and size $D(s\ge s_0)$ are shown in Fig.\ 3 for concentration
$c=0.5$, $\lambda =1$ (solid lines) and $\lambda =0.9$ (open circles).
The present results average over simulations for a long
time ($\approx 10^6 $ steps) from several configurations, each of
which is kept fixed for a particular simulation.  The slopes of the
straight lines in the case of conservative dynamics ($\lambda =1$)
determine the critical exponents according to $P(t) \sim
t^{-(1+\theta )}$ and $ D(s\ge s_0) \sim s^{-\tau }$.  We find the
best fit for $\theta =1.040 \pm 0.018$ and $\tau = 0.650 \pm
0.028$.  In this model the number of relaxed ``particles'' $n$ is not
proportional to the number of sites $s$ at which relaxation occurs,
thus leading to a new distribution $Q(n) \sim n^{-(1+\tau _n)}$.
We also calculated $\mathopen<n(l)\mathclose>$, to
obtain the exponent $\mu$ defined in Eq.~(\ref{newq}),
as well as the average number of topplings in a cluster of a
specified length $l$,
\begin{equation}
\mathopen< n \mathclose> _l  \sim l^{D_n}
\label{newq1}
\end{equation}
The scaling exponents are given in Table\ 1, and as we show in Section~5,
they satisfy various scaling relations to within numerical error.

It appears that the exponents are independent of the concentration
of defect bonds $c$, suggesting universal criticality.  Similar
robustness is exhibited by the system for variations in $\lambda $,
provided that $\lambda $ is strictly smaller than one:  namely, all
curves for $\lambda \mathopen< 1$ coincide with the open circle
curves in Fig.\ 3.

\section{Nonuniversal criticality: random-slope model~(C)}

Model C combines the critical height model with a critical slope
instability criterion at defect sites.  At these sites, which are randomly
distributed with relative concentration $c$, large columns of grains will
form; these relax according to an alternative set of rules: if at least
one of the two local slopes downstream of site $(i,j)$ exceeds a critical
value $\sigma _c$ then toppling occurs towards these neighbors.

\begin{equation}
{\mathrm{If}}~ \; \sigma _k(i,j) \equiv h(i,j) -h(i+1,j_k) \ge \sigma _c \, ,
\label{CSR}
\end{equation}
then
\begin{equation}
h(i,j) \to h(i,j) -1\, ; ~~~h(i+1,j_k) \to h(i+1,j_k) +1 \; , 
\label{CS2}
\end{equation}
where the index $k = \pm$ stands for right ($+$) or left ($-$) forward
neighboring site.  These rules are applied repeatedly until all slopes
become subcritical.  In order that topplings according to the critical
height rule, where we set $h_c =2$, do not affect those
topplings that proceed through critical slope dynamics, we
set $\sigma _c = 8 \gg h_c$.
After some transient time local slopes $\sigma (i,j) \approx \sigma _c -1$
are formed at randomly distributed defect sites, while the rest of the
system topples according to the critical height rule.  Due to the nonlocal
character of the critical slope rule in Eqs.\ (\ref{CSR}) - (\ref{CS2}),
topplings at defect sites may affect stability at their upstream neighbors
too, and these sites might then topple in the next time step.  In this way
an internal time scale is introduced, although the macroscopic transport
direction remains only one way, as in models A and B.

\begin{figure}[htb]
\epsfxsize=10cm
\epsffile[10 60 507 720]{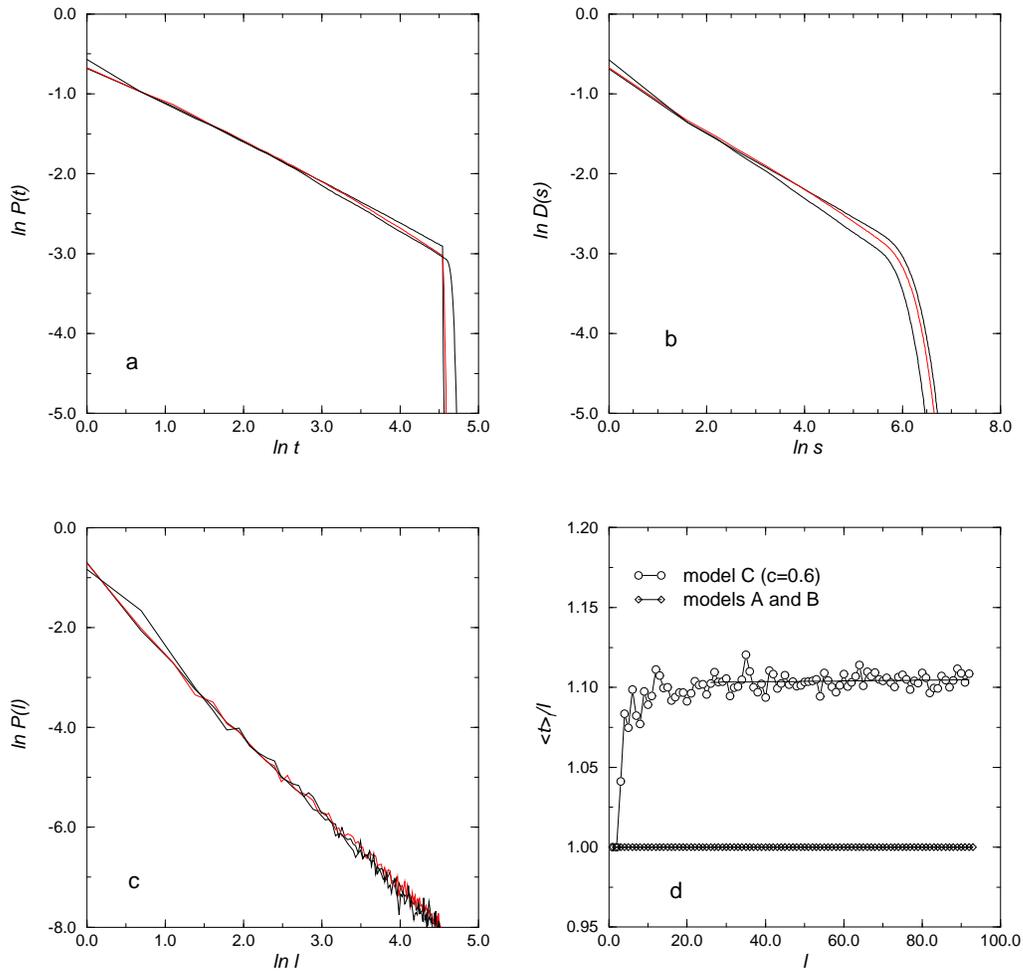}
\label{fig5}
\caption{Distributions of (a) duration, (b) size, and (c) length of 
the relaxation  clusters  for model C and (d) average duration of clusters 
of length $l$.}
\end{figure}

Results of the numerical simulations for the distributions of
duration $P(t \ge t_0)$, size $D(s\ge s_0)$, and length $P(l)$ are
shown in Fig.~\ref{fig5} for three concentrations of defect sites
$c = $ 0.05, 0.2, and 0.6.  The exponents for this model, $\theta $,
$\tau $, and $\alpha $ defined via $P(t\ge t_0) \sim t^{-\theta }$,
$D(s\ge s_0) \sim s^{-\tau }$, and $P(l)\sim l^{-(1+\alpha )}$
respectively, have a (weak) concentration dependence.  For instance,
$\theta = 0.516 \pm 0.002$ for $c=0.05$ increases to $0.581 \pm
0.001$ for $c=0.6$.  Similarly, $\alpha =0.499\pm 0.012$ at $c=0.05$
changes to $\alpha =0.563 \pm 0.024$ at $c=0.6$, and $\tau =0.354
\pm 0.001$ to $\tau =0.427 \pm 0.001$ in the same region.  For $c
\ge 0.7$ the curves exhibit a finite curvature due to the multiple
topplings.  (The exponents given in Table \ 1 are for $c=0.2$.)

The average duration of avalanches of selected length
$l$, $\mathopen<t\mathclose>_l$ in  Fig.~\ref{fig5}d, exhibits
scale invariance, in agreement with the scaling properties of 
the distributions, namely
\begin{equation}
\mathopen<t\mathclose>_l \sim l^z \, ,
\label{DEX}
\end{equation}

where $z$ is the dynamic exponent.  We find from Fig.~\ref{fig5}d 
that $z = 1.013 \pm 0.001$ for $c=0.6$.  The dynamic exponent
differs from 1 as a consequence of the nontrivial time scale in this
model, although the values of the other exponents are close to 
(but, as our numerical results suggest, distinct from) the
ones in the absence of defects.  However, nonuniversal properties
such as outflow current are different in the two cases:  there are
no apparent finite-size scaling effects in the presence of defects.
In the limit $c\to 1$, all sites become subject to the critical
slope rule, Eqs.~(\ref{CSR}) - (\ref{CS2}), due to which a finite slope is
formed.  SOC is lost, since every avalanche is of infinite duration:
the dynamics is dominated by single grain (one in/one out) events.

\section{Nonequilibrium phase transitions}

\begin{figure}[htb]
\epsfxsize=15cm
\epsffile[10 100 540 340]{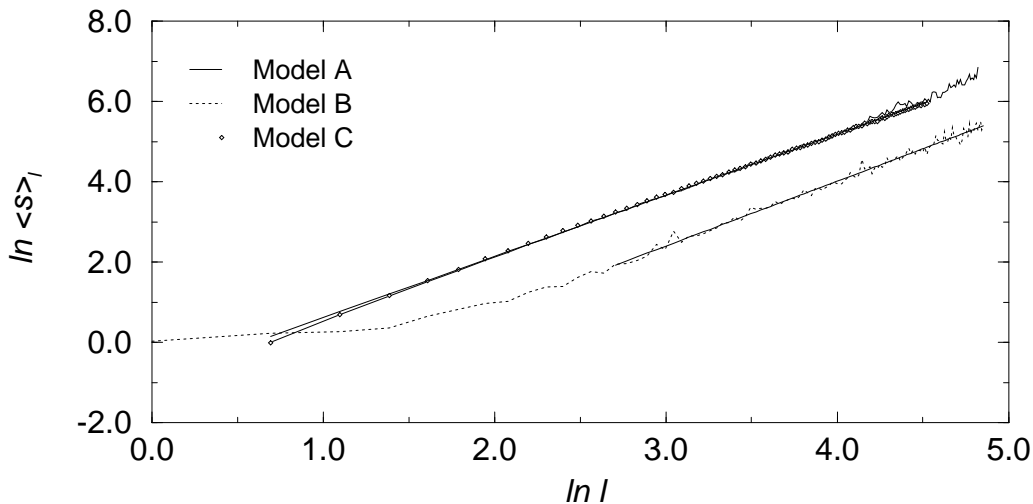}
\label{fig6}
\caption{Average size  of the relaxation clusters $\mathopen<s\mathclose>_l$
 of selected length $l$, measured parallel to the
transport direction, for  models A, B, and C, as indicated.}
\end{figure}

In the preceding sections we have shown that sandpile automata are
able to self-organize into a critical state in the presence of
frozen-in random defects, provided that the dynamics conserves the
number of grains at each time step.  The sets of critical exponents
for all three models are summarized in Table\ 1.  For model A
without defects the exponents are known exactly~\cite{DR}.  Model C
exponents are for concentration $c = 0.2$ of defect sites.  
(Note that in models A and  B time scale is measured in units of length,
therefore  the corresponding exponents are equivalent, {\it i.e.},
$\alpha \equiv \theta $.)
Also given
is the mass-to-scale ratio $D_{||}$ defined with respect to the
length parallel to the transport direction
\begin{equation}
\mathopen<s\mathclose>_l \sim l^{D_{||}} \, , \label{DFR}
\end{equation}

where $\mathopen<s\mathclose>_l$ is the time-averaged size of
clusters of selected length $l$.  In Fig.~\ref{fig6},
$\mathopen<s\mathclose>_l$ is obtained from separate numerical
simulations for models A, B, and C.  The exponent $D_n$ ({\it cf.}
Eq.(\ref{newq1})) is also given in Table \ 1.  The numerical values
of the exponents are in reasonable agreement with the following
scaling relation which can be shown to hold in all directed models:

\begin{equation}
 \theta \ z = D_{||}\ \tau = D_n\ \tau _n = \alpha \, .
\label{scal}
\end{equation}

\begin{center}
Table\ 1: Critical exponents and mass-to-scale ratio in 
2D directed models with defects
\vspace{1.5cm}
\begin{tabular}{|c|c|c|c|c|c|c|c|c|c|c|}\hline
Model& $D_{||}$ & $ \theta $ & $\tau $ & $\alpha $ & z & $\phi $& $\mu $
& $\tau _n $& $D_n$ &Remark \\ \hline
A & 3/2 & 1/2 & 1/3 & 1/2& 1& 1/2 & 1& 1/3& 3/2& universal  \\
\hline
B & 1.62& 1.04& 0.65 & 1.04& 1& none & 1.006& 0.466& 1.997& universal  \\
\hline
C & 1.49& 0.57 & 0.38& 0.54& 1.002& none & 0.978& 0.366& 1.54&  nonuniversal \\
\hline
\end{tabular}
\end{center}

An order parameter which is appropriate for the defect-mediated phase
transitions which occur in all three models can be defined as $q(c) = 1 -
\mathopen<J(c)\mathclose>/\mathopen<J(0)\mathclose>$, where
$\mathopen<J(c)\mathclose>$ the outflow current, which is the total number
of particles that flow over the lower boundary of the system in the
presence of defects of concentration $c$.  $\mathopen< \mathclose>$ denotes
an average over the total number of Monte Carlo steps.

\begin{figure}[htb]
\epsfxsize=15cm
\epsffile[20 100 564 400]{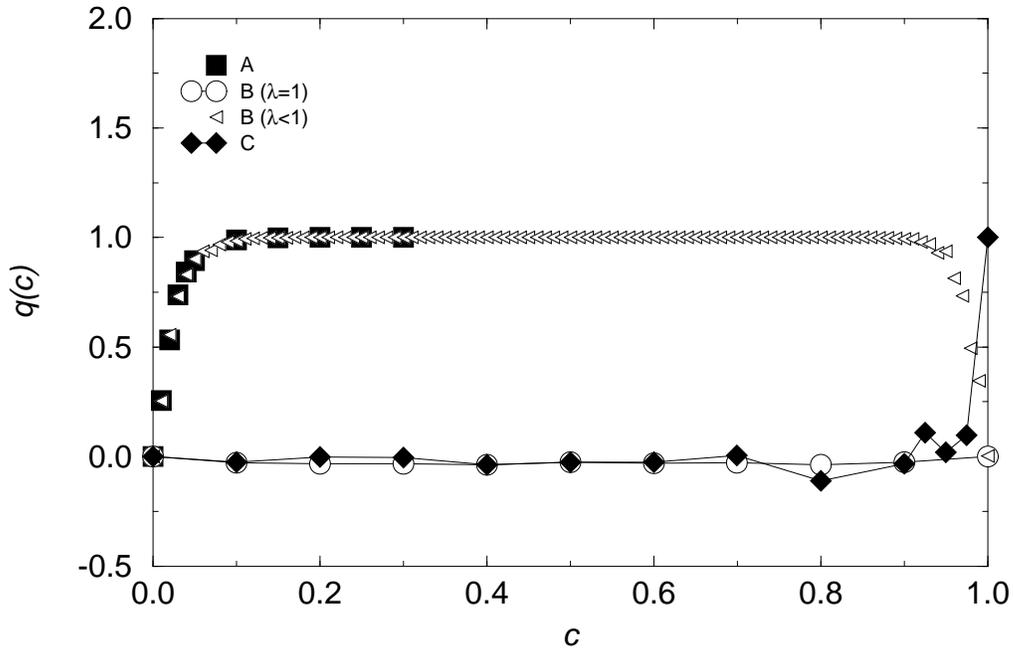}
\label{f7}
\caption{Order parameter $q(c)$ vs. 
concentration of defects $c$ for  models A, B, and C, as indicated.}
\end{figure}

In the models with conservative dynamics, {\it i.e.}, model B with
$\lambda =1$ and model C, the average flux out of the system is balanced
by the incoming current, thus leading to the vanishing of the order
parameter in the self-organized critical state, as shown in Fig.~6.  In
model A, however, $\mathopen<J(c)\mathclose>/\mathopen<J(0)\mathclose>$
decreases due to nonconservation of number of particles at defect sites,
leading to the appearance of finite order parameter $q$ with increasing
concentration of defect sites.  For  $c \ge c^*
= 0.295 \equiv 1-p_d$, the directed percolation threshold,
the states are such that only finite avalanches occur, while for $c
\mathopen < c^*$ there are
relaxation clusters of all sizes.  The slope of the
$q(c)$ curve at $c = c^*$ is close to zero.  In this way, model 
A is subcritical for finite
concentration $c$, which appears as a control parameter tuning the
coherence length of the self-organized state.  In the limit $c\to 0$ the
coherence length diverges, as discussed in Sec.\ 2.  Our numerical values
indicate that the order-parameter curve approaches the vertical axis with
a large but finite slope.  We find the same phenomenon in model B with
nonconserving dynamics ($\lambda \mathopen< 1$), where transfer is
incomplete at each negative bond.  (Model B is symmetric with respect to
transformation $c \to 1-c$, reflecting the symmetry between positive and
negative bonds, as can be seen in Fig.~6.)

There is another type of defect-mediated phase transition in model C
in the limit $c\to 1$. At this point SOC is lost in favor of a state
with finite net slope.  This nonequilibrium phase transition
requires a more detailed study.  In the related case of annealed
defects-- when the probability of toppling $p$ varies at each time step
but is the same for all sites in the system-- a collective phase
transition appears at $p_c = 0.293$~\cite{LTU}.

\section{Summary}

The presence of frozen-in defects in two-dimensional directed sandpile
automata leads to various new phenomena.  In this paper, we have
introduced and studied three variations of the directed abelian sandpile
automaton on the square lattice in order to explore this problem.  We find
that for site disorder, if the dynamics is nonconservative at defects, the
system is driven into a subcritical state with a finite correlation
length, which depends on the concentration $c$ of defects.  With any
concentration of bond disorder, the correlation length remains finite and
independent of the degree of nonconservation, provided that there is some
loss of conservation, $\lambda \mathopen< 1$.  (Other models of
nonconservative cellular automata have been studied
recently~\cite{noncon}, and are seen to have robust SOC behavior.  Here we
have lack of conservation occurring {\it solely} due to the presence of
defects.) Furthermore, if the dynamics is conservative, the automaton
self-organizes into a critical state with universal scaling properties.
For a third type, the case of random slope defects, which correspond most
closely to the random field version of analogous spin problems, we observe
some indications of nonuniversal, concentration-dependent scaling
exponents.  Scaling relations between the exponents are fulfilled exactly
for the first model (with $c = 0$), and within numerical error for the
latter two models for each concentration of disorder.  Finally, varying
the concentration of defects appears to be a mechanism for continuously
tuning the local rules of relaxation, which may eventually lead to a phase
transition between metastable states with different properties.

\noindent


\section*{Acknowledgments} This work was supported by the
Ministry of Science and Technology of the Republic of Slovenia. We thank
R. Pirc, J. Kert\'esz and P. Grassberger for critical comments on the
manuscript.



\begin{thebibliography}{99}

\bibitem{BTW} P. Bak, C. Tang and K. Wiesenfeld, Phys.Rev.Lett.  {\bf 59}, 381 (1987). \\   
              P. Bak, C. Tang and K. Wiesenfeld, Phys.Rev. A {\bf 38}, 364 (1988).

\bibitem{DR} D. Dhar and R. Ramaswamy, Phys. Rev. Lett. {\bf 63}, 1659 (1989).            

\bibitem{MKK} S.S. Manna, L.B. Kiss and J. Kert\'esz, J. Stat. Phys. {\bf
61}, 923 (1990).

\bibitem{TNURP} B. Tadi\'c, U. Nowak, K.D. Usadel, R. Ramaswamy, and 
                S. Padlewski,  Phys. Rev. A {\bf 45}, 8536 (1992).

\bibitem{Theiler} J. Theiler, Phys. Rev. E {\bf 47}, 733 (1993).

\bibitem{dissip} The number of particles relaxed up to distance
$l$ is the total number of particles that take part in the relaxation 
processes from the top of pile up to  the $l$'th row. For
lattices with defects, this quantity need not be the same as the total
number of toppled sites.

\bibitem{dirperc} J.W. Essam, A.J. Guttmann, and K. de'Bell, J. Phys.
                 A {\bf 21}, 3815 (1988).

\bibitem{Ketal} L.P. Kadanoff, S.R. Nagel, L.Wu and S.M. Zhou, Phys.Rev. A  
                {\bf 39}, 6524 (1989).	

\bibitem{MH} E.N. Miranda and H.J. Herrmann, Physica A {\bf 175}, 339 (1991);
B. Tadi\'c, J. Non-Cryst. Solids {\bf 172-174}, 501 (1994).

\bibitem{Peng} G. Peng, Physica A {\bf 201}, 573 (1993).

\bibitem{LTU} S. L\"ubeck, B. Tadi\'c, and  K.D. Usadel, in {\it Fractals 
in the Natural and Applied Sciences}, edited by M.M. Novak, Chapman and Hall
(1995), and to be published.

\bibitem{noncon} See {\it e.g.}, K.  Christensen and Z.  Olami,
Phys.  Rev.  A {\bf 46}, 1829 (1992); A.  Corral, C.J.  P\'erez, A.
D\'\i az-Guilera, and A.  Arenas, Phys.  Rev.  Lett.  {\bf 74}, 118
(1995).

\end{thebibliography}
\end{document}